\documentstyle[12pt,preprint]{aastex}
\shorttitle{Warped bar of LMC}
\shortauthors{A. Subramaniam}
\begin{document}
\title{LMC Bar: evidence for a warped bar}
\author{Annapurni Subramaniam\altaffilmark{1}}
\affil{Indian Institute of Astrophysics, Koramangala II Block, Bangalore - 34}
\begin{abstract}
The geometry of the LMC bar is studied using the de-reddened mean
magnitudes of the red clump stars ($I_0$ ) from the OGLE II catalogue. 
The value of $I_0$ is found to vary in the east-west direction such that both the east and the 
west ends of the bar are closer to us with respect to the center of the bar. 
The maximum observed variation has a statistical significance of more than 7.6 $\sigma$ with
respect to the maximum value of random error. The variation in $I_0$ indicates the presence
of warp in the bar of LMC.
The warp and the structures
seen in the bar indicate that the bar could be a dynamically disturbed structure.
\end{abstract}
\section{Introduction}
The Magellanic Clouds and Milky Way are known to have experienced close encounters.
These encounters  would result 
in tidal forces which could alter the structure of the Clouds.
\citet{vc01}, \citet{v01}, \citet{wn01}, have studied the geometry 
of the LMC using DENIS and 2MASS data, and found evidences for tidal signatures in the LMC. 
These studies were done on the outer regions of the LMC, 
at radial distances more the 3 degrees. The bar region of the LMC has not been 
studied in detail, from the geometry and structure point of view, though it was 
covered in the study of the total structure of LMC by \citet{v01}. \citet{v01} 
found that the density structure of the bar region is very smooth and no features 
were detected. The interesting point which they noticed, but to which they did not give 
much importance was the change in the major axis position angle 
within a radial distance of 3 degrees. 

\citet{os02} studied the LMC outer regions 
and found evidences for a possible 
warp in the south-west of the LMC and argued that the LMC plane is warped and twisted,
containing features that extend up to 2.5 Kpc out of the plane. The warp as found by 
\citet{os02} could have started closer to the LMC center and it will be interesting 
to find the starting point of this deviation. There has been a lot of recent photometric 
surveys and OGLE II \citep{u00} survey covers most of the bar region and thus it is well
 suited for this study. We used the brightness of core helium-burning red clump stars
 in the bar region of the LMC as a probe for the bar structure. 
The difference in the de-reddened mean magnitude of the red clump stars is used as 
differential distance indicator. The technique used here is identical to the one used 
by \citet{os02}. We look for evidences of tidal interaction in the bar region, 
like the presence of a warp, within a radial distance of 3 degrees. 

\section{Data selection and estimation of de-reddened mean magnitude ($I_0$)}
OGLE II survey \citep{u00} consists of photometric data of 7 million stars in B, V and I 
pass bands in the central 5.7 square degree of LMC. The data is presented for 21 regions,
 which are located within 2.5 degree from the optical center of the LMC.
 Initially, the total observed region is divided into 336 sections of size 7.1$\times$7.1
 arcmin$^2$ each. The red clump stars are identified using I vs (V$-$I) colour-magnitude
diagram (CMD) and 
on an average, 7000 red clump stars were identified per region.
 
\begin{figure}
\plotone{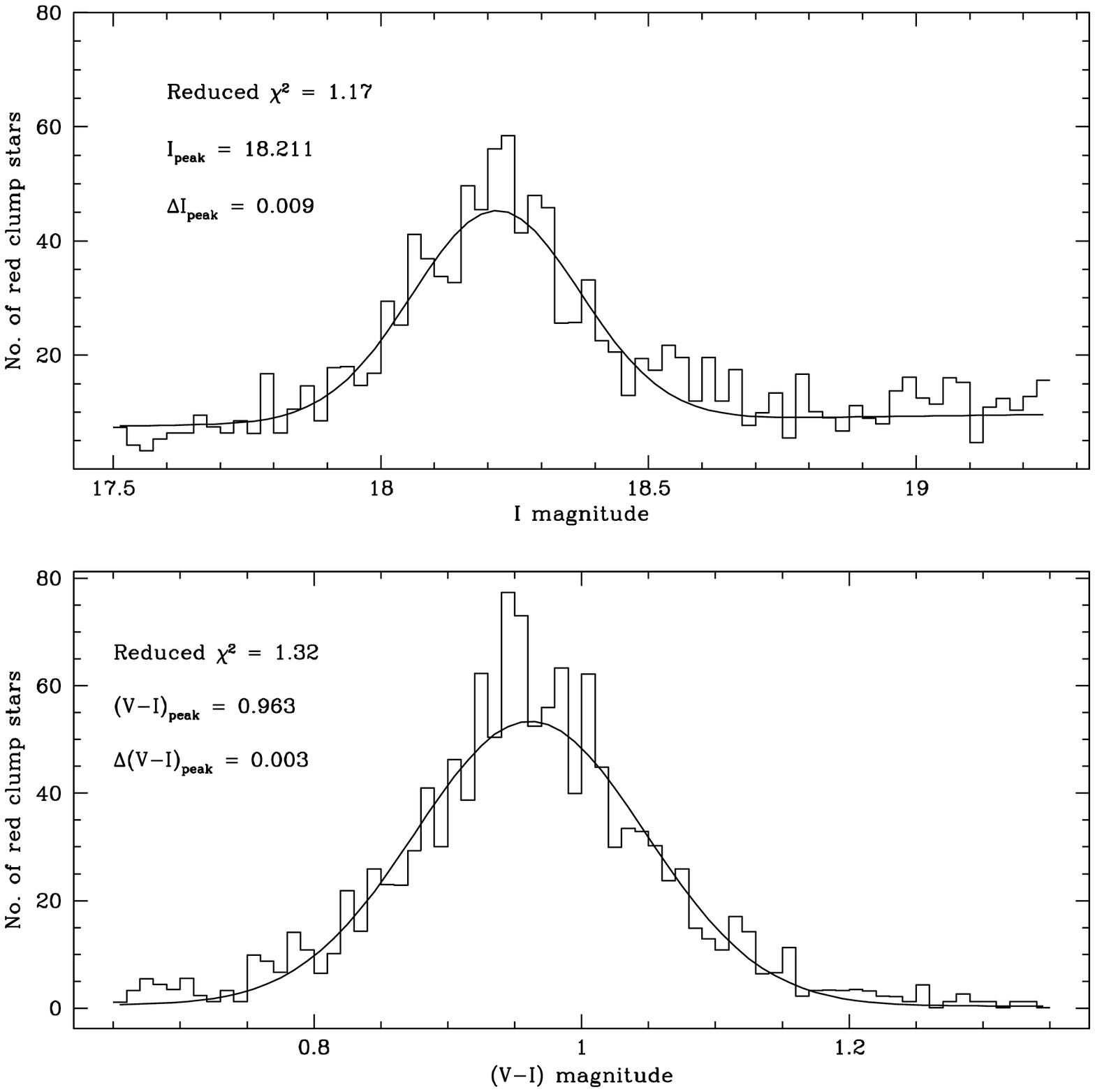}
\caption{ The luminosity function of red clump stars are plotted against I magnitude
and (V$-$I) colour. The reduced $\chi^2$ value of the fit of the function to the distribution,
the estimated values of the peak and its error are indicated in the figure. This data is
obtained within a area of 3.56$\times$3.56 arcmin$^2$ centered at  
the location RA = 5$^h$ 32$^m$ 45$^s$ and Dec = $-$70$^o$ 25' 37''. 
\label{fig1}}
\end{figure}

The data suffers from the incompleteness problem due to crowding effects, and 
the incompleteness in the data in I and V pass bands are tabulated in \citet{u00}.
The frequency distribution of red clump stars in each CMD is estimated in both I magnitude 
and (V$-$I) colour after correcting for the data incompleteness, using
a bin size of 0.015 mag for (V$-$I) colour and 0.025 mag in I magnitude. These 
distributions are fitted with the Gaussian+ a quadratic function, similar to \citet{os02}.
 The distributions are fitted using a non-linear least square fits, to obtain the best 
fitting parameters when the $\chi^2$ value is minimum. The parameters estimated are 
the peak of the function, error in the estimated peak value, the width of the profile 
and the goodness of fit. The above 
described area is chosen so as to have a good number of stars for the fit. On the other 
hand, the choice of area should not affect the conclusions derived here. Therefore, 
two more data sets were created by dividing the observed area into 672 sections 
(3.56$\times$7.1 arcmin$^2$) and 1344 sections (3.56$\times$3.56 arcmin$^2$),
 and the frequency distribution and the fit parameters were estimated for 
these data also. The number of red clump stars in these area bin also scale like 
the area. 
In order to estimate the goodness of fit, the reduced $\chi^2$ values were estimated.
After rejecting the distributions with very high $\chi^2$ values, the 
average values of reduced $\chi^2$ for the fit of the I mag distribution
 are 1.68, 1.60 and 1.59
 and those for the (V$-$I) distribution are 1.65, 1.49 and 1.39, for the largest,
medium and smallest area bin respectively. It can be seen that the choice of area
 does not affect the shape of the distribution very much and hence the derived parameters. As the 
average value of the reduced $\chi^2$ is found to be minimum for the smallest area bin, 
smallest area bins are chosen for further analysis.
A typical frequency distribution of red clump stars in I and (V$-$I) magnitudes
are shown in figure~\ref{fig1}. The reduced $\chi^2$ value of the fit, estimated value of the peak and its error
are also indicated in the figure. For further analysis, we restrict the data points with 
reduced $\chi^2$ value less than 2.6, which reduces the number of regions to 1191.

The peak values of the colour, (V$-$I) mag at each location is used to estimate the 
reddening. The reddening is calculated using the relation
 E(V$-$I) = (V$-$I)$_(obs)$ -- 0.92 mag.
The intrinsic colour of the red clump stars is assumed to be 0.92 mag \citep{os02}.
The interstellar extinction is estimated by $A_I$ = 1.4\,E(V$-$I) \citep{sfd98}.
After correcting the mean I mag for interstellar extinction, $I_0$ for each region 
is estimated. 

\begin{figure}
\plotone{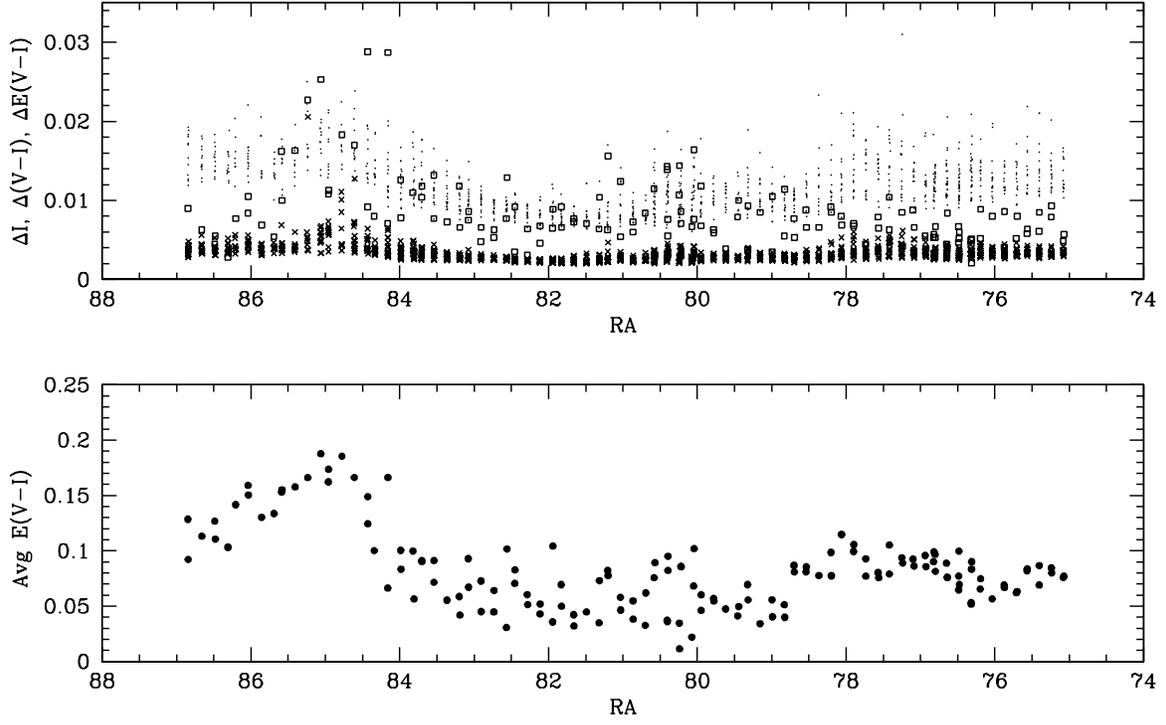}
\caption{{\it Top panel:} The error in the estimation of the peak of 
I magnitude (dots) and (V$-$I) colour
(crosses) of the red clump luminosity function are shown as a function of RA for
1191 regions. The
dispersion in the reddening E(V$-$I) is indicated as open squares. 
{\it Bottom panel:} The average reddening E(V$-$I) is shown as a function of RA.
\label{fig2}}
\end{figure}

The error in the estimation of the peak values of the I and (V$-$I) distribution are 
shown as a function of RA in figure~\ref{fig2}. 
Since the data spans about 12 degree in RA and the span in declination is only 
about 2 degrees,
the error variations are shown as a function of RA. 
In general, most of the data points have
$\Delta (V-I)_{peak} \le 0.005$ mag and $\Delta I_{peak} \le 0.02$ mag and
about 2\% of the points have slightly higher errors.
We have chosen regions which have errors less than or equal to the values indicated above,
for further analysis.
Since the random error in the estimation of $I_0$ has contribution from $\Delta (V-I)_{peak}$
and $\Delta I_{peak}$, then $\Delta I_0$ can be estimated as, 
$\Delta I^2_0 = ((\Delta (V-I)_{peak}^2)+(\Delta I_{peak}^2)$.
If we consider the maximum errors in the peak values, which are 
$\Delta (V-I)_{peak} = 0.005$ and $\Delta I_{peak} = 0.02$,
 then the maximum random error in $I_0$ is $\Delta_{max} I_0  = 0.021$ mag.
Another factor which can contribute to the total error is the reddening.
The E(V$-$I) reddening values of 8 adjacent regions were averaged and the standard 
deviation in the mean reddening was estimated. This value of standard deviation in the 
reddening is also plotted in figure~\ref{fig2}. 
The average reddening is found to be E(V$-$I) = 0.081 $\pm$ 0.004 mag and these values as a
function of RA is shown in the bottom panel of figure~\ref{fig2}.
It can be seen that $\Delta_{max} E(V-I) = 0.029$ mag. The
rejection criteria based on the errors in the peak value, as mentioned above
  removes the regions which show
higher deviation in reddening also. Hence after the rejections, 
the $\Delta_{max} E(V-I)$ is found to be 0.02 mag.
A systematic error in the estimation of $I_0$ can arise from $\Delta E(V-I)$.
If one considers 0.004 mag as the standard deviation in reddening estimate, then
the total error due to random as well as systematic effects is 0.021 mag, which is
same as $\Delta_{max} I_0$.
If we take a value of 0.02 mag as the upper limit in the standard deviation in 
reddening, then the maximum total error due to all the three sources, is 0.029 mag.
The zero-point error and photometric errors 
are not included here, as they are almost the same for the entire data, the data 
being homogeneous. 

\section{Results and Discussion}
\begin{figure}
\plotone{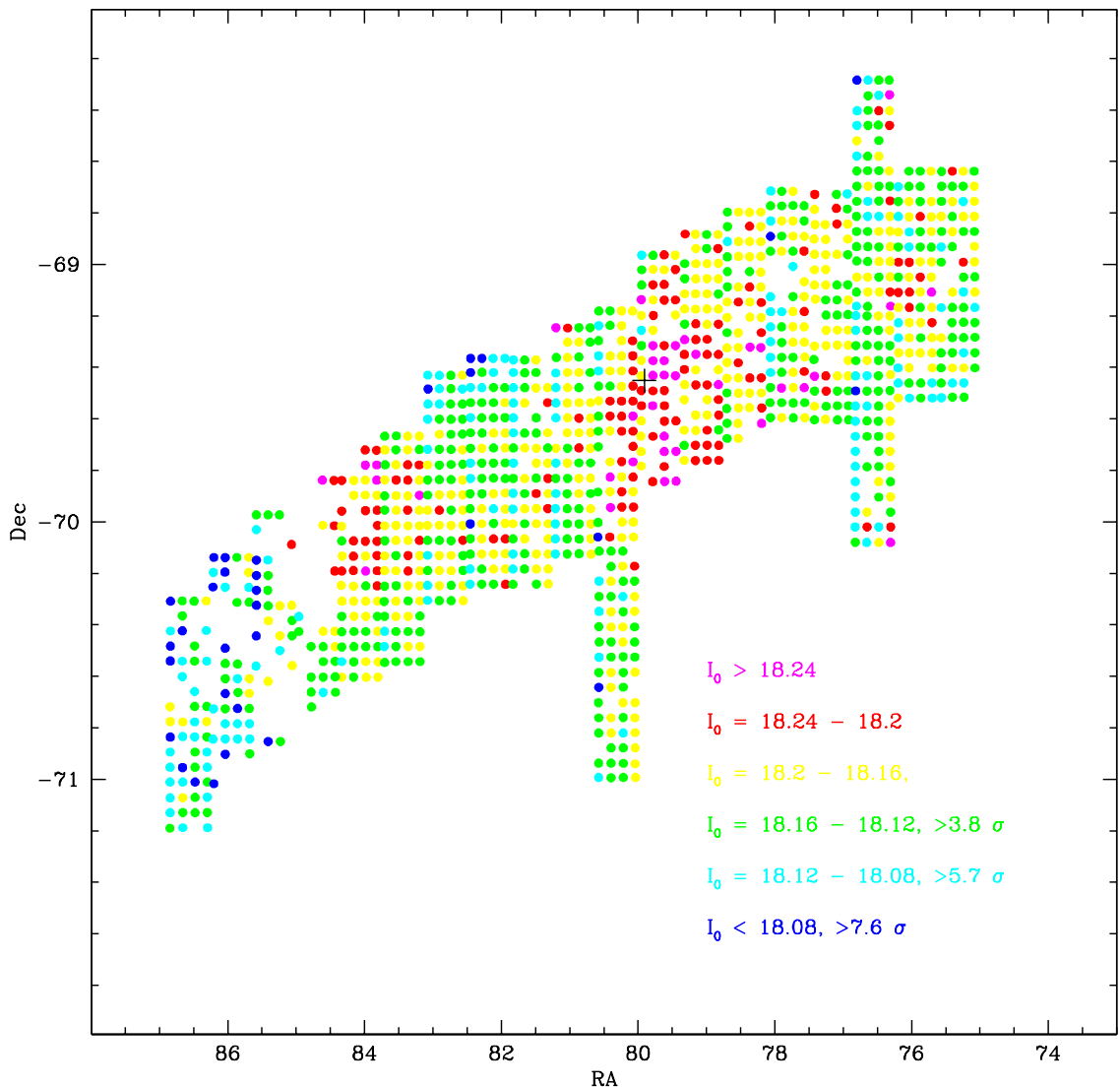}
\caption{A 2-D plot of the de-reddened mean red clump magnitude, $I_0$, for
1123 regions are shown.
The colour code used is such that the colour magenta indicates location which are farthest
and blue indicates regions which are closest. The statistical significance of the $I_0$ bins
with respect to maximum random error is also indicated in the figure. The black plus sign
shows the location of the LMC center.
\label{fig3}}
\end{figure}

The 2D figure of the region studied is shown in figure~\ref{fig3}, where the variation in $I_0$is
shown as a function of RA and Dec.  The farthest points have $I_0$ more than 18.24 mag
and the closest points have $I_0$ less than 18.08 mag. This corresponds to a net difference
of more than 0.16 mag. This value is more than 7.6 times the maximum random
error and more than 5.6 times the maximum total error. Hence the net variation in the 
de-reddened mean red clump magnitudes is statistically significant.

At locations RA= 79$^o$.5 and Dec = $-69^o$.6 and RA= 84$^o$.5 and Dec=$-70^o$, 
the $I_0$ values
are higher indicating that these regions are located at a larger distance.
The regions in between the above points are closer to us. The eastern most regions
are closest, as indicated in the figure. At RA=84$^o$.5,
another feature which can be noticed is that along the declination axis, there is a
change in the relative distance. This is such that the northern regions are farther and
the southern regions are closer. The difference in $I_0$ is more than 
3.8 times the $\Delta_{max} I_0$.
The center of the LMC is taken to be $05^h19^m38^s.0$ $-69^o27'5''.2$ (2000.0)
\citep{df73}. Then the center lies near the fainter $I_0$ points located around RA=79$^o$.5.
Thus the regions westward of the center are also found to be closer.

\begin{figure}
\plotone{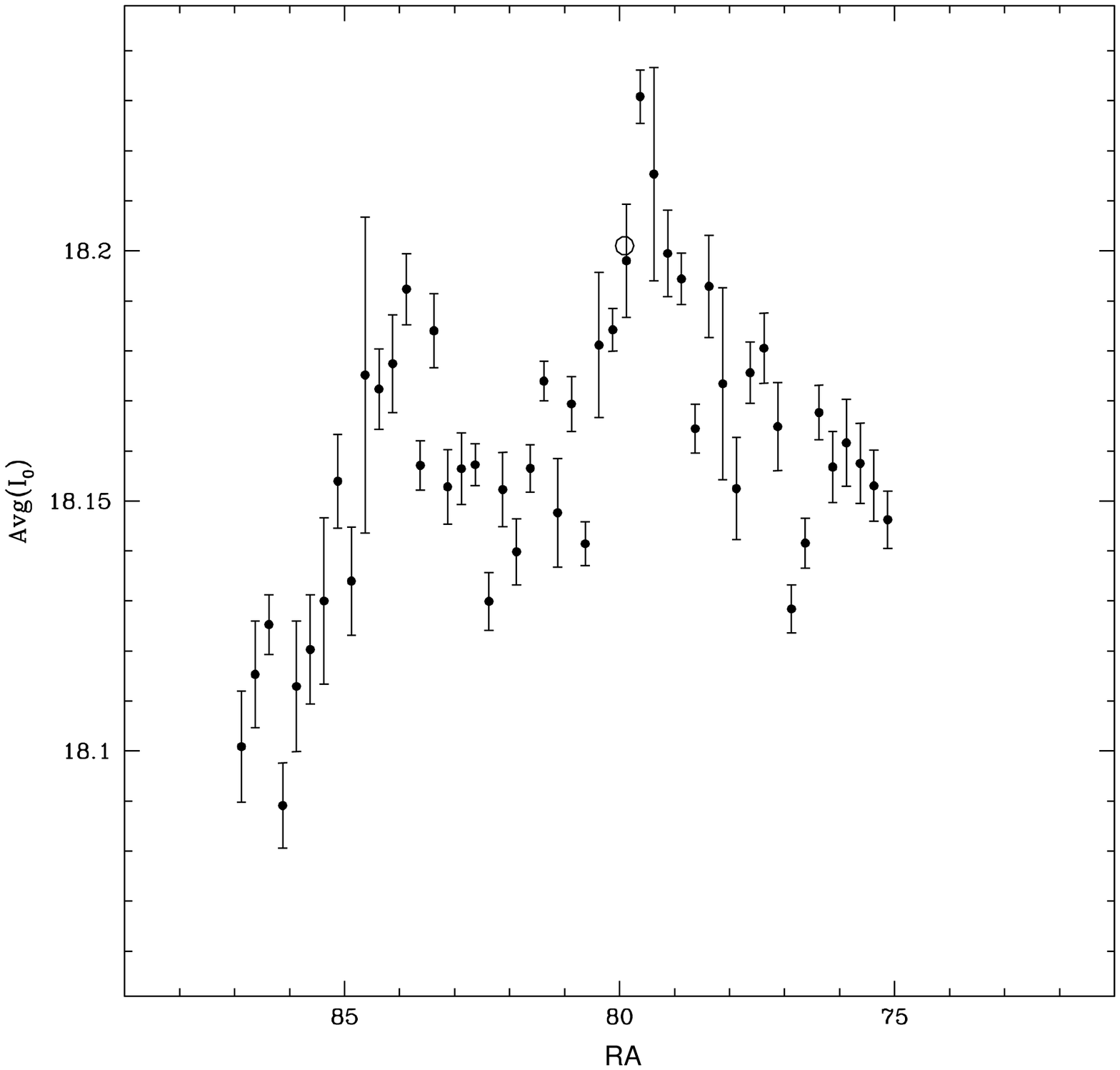}
\caption{The average of the $I_0$ along the declination is estimated for various RA and
the value avg($I_0$) is shown against RA. The error indicates the scatter in $I_0$
along the declination. The open circle shows the location of the center of LMC.
\label{fig4}}
\end{figure}

In order to study the variation of $I_0$ along RA, $I_0$ values along declination
are averaged, and a plot of avg($I_0$) versus RA is shown in figure~\ref{fig4}. The error bars indicate
the deviation in $I_0$ along the declination, for a given RA. It can be seen that
there are variations in the $I_0$ magnitude along the bar. The center of LMC is shown as
an open circle. Most striking feature is the wavy pattern in $I_0$. 
The eastern side of the bar is closer
to us, when compared to the bar region near the center. So also is the western side closer
to us. We see an M-type variation in $I_0$ along the RA. Thus the features indicate that the
bar of the LMC is warped. 
It would be interesting to find the relative
inclinations of the disk and the bar, as this will
help us estimate their locations.  The geometry of the bar based on the $I_0$ variation
will be presented in another paper, which is in preparation.
It would be ideal to use the photometric data of the LMC stars from other surveys, 
like, MACHO survey to estimate the disk parameters.

The structure of the bar as derived here is delineated by stars belonging to the
red clump population. The techniques used in this study are used earlier by many 
studies \citep{v01,u98,vc01}. The data used here have been used by \citet{u98} 
to estimate distance to the LMC, but they have not used it to study the relative 
distances within LMC. The reddening is found to be almost a constant along the bar, except
in the east end, therefore the  magnitude variation is not an imprint of reddening  in the bar.
The variation in the red clump luminosity could also be due to the age and metallicity
difference, rather than due to the relative distance. The study by \citet{sa02}
on the local stellar population of nova regions found no major difference in the 
population of the intermediate age stars in the Bar region. This is again supported 
by the findings of \citet{os02} and \citet{vc01}. Hence the variation 
seen in $I_0$ magnitude is mainly due to the geometry of the bar. 
The estimates of self-lensing optical depth in LMC appear to be too low to account fully for the
entire microlensing optical depth \citep{g95}.
This kind of a structure in the bar would contribute to more optical depth within LMC,
which can increase the self lensing within LMC.

The LMC bar is thus found to show structures. The presence of
warp in the bar indicates that the bar is dynamically disturbed. Since the bar is located
well within the tidal radius of LMC, the tidal effects due to LMC-SMC-Galaxy interaction may
not be the cause of the disturbance. On the other hand, if the bar is not aligned 
with the disk, then the disk can induce perturbations on the bar. This in turn can 
create structures in the bar. In order to explain the LMC microlensing events, 
\citet{ze00} proposed the bar to be an unvirialised structure, which is slightly mis-aligned 
with and offset from the LMC disk. \citet{ze00} also claimed that the interactions of the 
Magellanic Clouds with the Galaxy could be responsible for the misalignments and displacements 
of the bar with respect to the disk. Therefore, it is necessary to
find out the the geometry of the bar and the disk, which would throw light on the
source of perturbations in the bar. 

I thank Ram Sagar, Uma Gorti, Anupama and Kiran Jain for helpful discussions.
I also thank the referee for his comments which improved the presentation of the paper.

\clearpage
\clearpage

\clearpage

\clearpage
\end{document}